\documentclass[iop]{emulateapj}
\usepackage{color} 

\newcommand{\gke}{{\tt Gkeyll}}

\slugcomment{(Received December 16, 2018; Revised February 5, 2019; Accepted February 6, 2019 )}
\shorttitle{}
\shortauthors{V. Skoutnev, A. Hakim, J. Juno, J. TenBarge}

\begin{document}


\title{\sc{Temperature-dependent Saturation of Weibel-type Instabilities in Counter-streaming Plasmas}}


\author{V. Skoutnev\altaffilmark{1,2}}
\email{skoutnev@princeton.edu}

\author{A. Hakim\altaffilmark{2}}
\author{J. Juno\altaffilmark{3}}
\author{J. M. TenBarge\altaffilmark{1}}







\altaffiltext{1}{Department of Astrophysical Sciences, Princeton University, Princeton, NJ 08544, USA}
\altaffiltext{2}{Princeton Plasma Physics Laboratory, Princeton, NJ 08543, USA}
\altaffiltext{3}{IREAP, University of Maryland, College Park, MD 20742, USA}


\begin{abstract}
We present the first 2X2V continuum Vlasov-Maxwell simulations of interpenetrating, unmagnetized plasmas to study the competition between two-stream, Oblique, and filamentation  modes in the weakly relativistic regime. We find that after nonlinear saturation of the fastest-growing two-stream and Oblique modes, the effective temperature anisotropy, which drives current filament formation via the secular Weibel instability, has a strong dependence on the internal temperature of the counter-streaming plasmas. The effective temperature anisotropy is significantly more reduced in colder than in hotter plasmas, leading to orders of magnitude lower magnetization for colder plasmas. A strong dependence of the energy conversion efficiency of Weibel-type instabilities on internal beam temperature has implications for determining their contribution to the observed magnetization of many astrophysical and laboratory plasmas. 

\end{abstract}




\section{Introduction}
Electromagnetic instabilities of the Weibel-type have been extensively studied as mechanisms for generating the sub-equipartion, long-lived magnetic fields whose effects are inferred or observed in laser-driven plasmas \citep{fox13,hunt15}, gamma ray bursts \citep{medved99}, pulsar wind outflows \citep{kazimura98}, and cosmological scenarios \citep{schlick03,lazar09}. Interpenetrating, unmagnetized plasma flows common to the initial conditions of these systems support a diversity of competing instabilities spanned by electrostatic two-stream (TS) and electromagnetic Oblique and filamentation modes \citep{bret09}. Obtaining the efficiency $\epsilon_B=E_M/E_K$ of the conversion of initial kinetic energy $E_K= \langle\frac{1}{2}\rho v^2\rangle$ into magnetic energy $E_M=\langle B^2/2 \mu_0\rangle$ by all of the simultaneously present instabilities is critical for estimating the role that Weibel-type generated magnetic fields play in both laboratory and astrophysical contexts. 

Magnetic field generation in counter-streaming plasmas is associated with the filamentation instability (FI; \cite{fried59}), while in plasmas with anisotropic bi-Maxwellian velocity distributions, it is attributed to the Weibel instability (WI; \cite{weibel59}). However, the secular WI can also be present in counter-streaming plasmas after saturation of a faster growing TS mode relaxes the initial plasma velocity distribution to an anisotropic bi-Maxwellian \citep{schlick03}. In this Letter, we refer to these modes as Weibel-type instabilities, e.g. the FI and WI.

All of these modes grow on plasma frequency time scales and therefore require kinetic numerical simulations to study their evolution and competition in the nonlinear regime. Both subrelativistic and relativistic PIC simulations in 2X and 3X \citep{fonseca03,silva03,sakai04, nishi03,nishi05,kato08,takamoto18} have shown promising efficiencies of $ 0.25\%\lesssim  \epsilon_B\lesssim 2\%$ in the long-term quasi-steady state. Typical simulations proceed first forming electron Alfv\'{e}n limited current filaments on small scales that coalesce into larger filaments, followed by a similar process for ions \citep{fred04,hededal04}. We note that recent 2X PIC simulations have found the merging process to disrupt the coherent filament structures, reducing $\epsilon_B$ \citep{kumar15,takamoto18}. Filament merging and break-up events along with various electron trapping and acceleration mechanisms have been found to heat electrons, possibly explaining the origin of nonthermal radiation observed in several astrophysical scenarios.

Continuum Vlasov-Maxwell codes are an alternative approach that has historically been lucrative in the 1X2V setting \citep{calif98}, confirming magnetic trapping as the mechanism responsible for saturation \citep{davidson72} and finding $\epsilon_B \sim 1\%$ for weakly relativistic drift velocities $0.1 < u_d/c < 0.6$. However, continuum Vlasov-Maxwell methods have not been employed as widely as the PIC method due to the expense in directly discretizing the distribution function on a high dimensional phase space grid. Nevertheless, advancements in algorithms and computing power have led to a resurgence in the application of continuum Vlasov-Maxwell codes to problems requiring kinetic modeling, motivated by the advantages of the direct-discretization noise-free representation of the distribution function. For example, Cagas et al. (2017) found that for colder beams, electrostatic potentials due to secondary streaming induced by the FI play an equally dominant role as magnetic trapping in saturation of the 1X FI. 

Using the same method as Cagas et al. (2017), this Letter presents the first 2X2V continuum Vlasov-Maxwell simulations of interpenetrating plasma flows aimed at incorporating the competition of the two-stream and Oblique modes with the FI. We find that the magnetization parameter $\epsilon_B$ is strongly dependent on the internal temperature of the beams in the weakly relativistic regime, wherein the growth rates of the various modes are comparable. Saturation of TS and Oblique modes in colder beams tends to scatter the anisotropy in the particle velocity distribution that drives the later formation of the current filaments via the secular WI. This behavior results in a complete lack of magnetization, $\epsilon_B\lesssim 10^{-5}$, in the nonlinear saturation phase of colder beams and typically reported values of $\epsilon_B\sim 10^{-2}$ for hotter beams.

\section{Problem Setup}
We work in the framework of the Vlasov-Maxwell system. The Vlasov equation for species $s$ is given by
\begin{equation}
\frac{\partial f_s}{\partial t}+\vec{v}\cdot \frac{\partial f_s}{\partial {\vec{x}}}+\frac{q_s}{m_s}(\vec{E}+\vec{v}\times \vec{B})\cdot\frac{\partial f_s}{\partial{\vec{v}}}=0,
\end{equation}
where $f_s$=$f_s(x,y,v_x,v_y,t)$ is the species distribution function and $\vec{E}=E_x(x,y)\hat{x}+E_y(x,y)\hat{y}$ and $\vec{B}=B_z(x,y)\hat{z}$ are the electric and magnetic fields that are evolved using Maxwell's equations. 

The initial plasma flow is modeled as two uniform density electron streams with opposite flow speeds $u_d$ in the y direction, allowing study of the purely reactive ($\omega=\mathrm{Im}[\omega]>0$) form of the FI. The beams are neutralized by a uniform background of ions. The distribution function at $t=0$ for electrons is given by
\begin{equation}
f_{0,e}(v_x,v_y)=\frac{1}{2\pi v_{th}^2}e^{-\frac{v_x^2}{2v_{th}^2}}\left[e^{-\frac{(v_y-u_d)^2}{2v_{th}^2}}+e^{-\frac{(v_y+u_d)^2}{2v_{th}^2}}\right],
\end{equation}
where $v_{th} = \sqrt{k_B T_e/m_e}$.

The counter-streaming motion provides the free energy source that drives the TS, Oblique, and filamentation instabilities, which convert kinetic energy into electric and magnetic energy. TS and FI modes generate primarily electrostatic and magnetic energy, respectively, while Oblique modes qualitatively develop as a mixture of the two instabilities.  

To study this problem in the nonlinear regime, we perform continuum kinetic simulations in two periodic configuration space dimensions and two velocity dimensions (2X2V) using the \gke~framework \citep{juno18}. \gke~uses the discontinuous Galerkin method for spatial discretization and a strong-stability preserving Runge-Kutta method for time stepping to solve the full Vlasov-Maxwell system in the non-relativistic regime.

\section{Linear Theory}
We define the angle $\theta$ of the wave vector $\vec{k}$ from the x-axis so that $\theta=0^{\circ}$ corresponds to pure FI modes, $\theta=90^{\circ}$ to pure TS modes, and any angle in between to Oblique modes. Linearizing the Vlasov-Maxwell system gives the general dispersion matrix 
\begin{equation}
D_{ij}=\frac{\omega^2}{c^2}\left( k_ik_j-k^2\delta_{ij}\right) +\epsilon_{ij},
\end{equation}
with
\begin{equation}
\epsilon_{ij}=\left(1-\sum_s\frac{\omega_{p,s}^2}{\omega^2}\right)\delta_{ij}+\sum_s \frac{\omega_{p,s}^2}{\omega^2}\int_{-\infty}^{\infty} v_iv_j\frac{\vec{k}\cdot \nabla_{\vec{v}} f_{0,s}}{\omega-\vec{k}\cdot \vec{v}}d^2v.
\end{equation}

\begin{figure}[h]
\includegraphics[width=\linewidth]{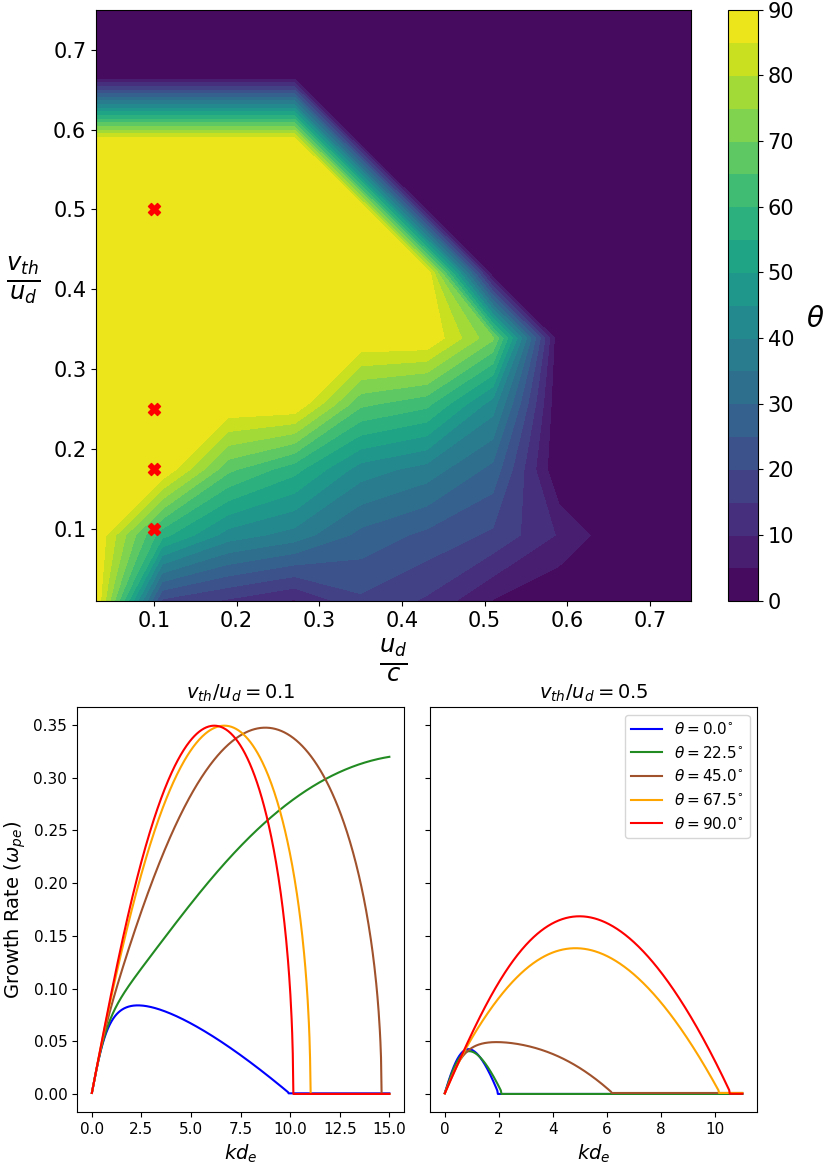}

\caption{ Top panel: contour plot of the angle of the fastest-growing mode in the parameter space of $v_{th}/u_d$ and $u_d/c$. $\theta=90^{\circ}$ corresponds to pure TS and $\theta=0^{\circ}$ to pure FI modes. Red crosses correspond to the four simulations presented. Bottom panels: growth rates versus wavenumber of different modes for the hot (right panel) and cold (left panel) cases at $u_d=0.1c$. }\label{fig:angle_growth}

\end{figure}

We find that using $\vec{k}$ aligned coordinates gives the simplest representation of $D$, i.e., rotating the original coordinates by angle $\theta$. Substitution of $f_{0,e}$ into the general dispersion matrix and neglecting the ion contribution then gives 
\begin{equation}\label{eq:D}
D=\left(
\begin{array}{cc}
D_{11}  & D_{12}\\
 D_{21}&D_{22} 
\end{array}\right)
\end{equation}
$$D_{11}=1-\frac{\omega_{pe}^2}{4k^2v_{th}^2}\left(Z'(\xi_{+})+ Z'(\xi_{-})\right)$$
$$D_{12}=D_{21}=\frac{\omega_{pe}^2 u_d \cos\theta}{4\omega kv_{th}^2}\left(Z'(\xi_{+})- Z'(\xi_{-})\right)$$
$$D_{22}=1-\frac{\omega_{pe}^2}{\omega^2}-\frac{k^2c^2}{\omega^2}- \frac{\omega_{pe}^2\left(u_d^2 \cos^2\theta+v_{th}^2\right)}{4\omega^2 v_{th}^2}\left(Z'(\xi_{+})+ Z'(\xi_{-})\right),$$
where $Z(\xi_{\pm})$ is the plasma dispersion function with $\xi_{\pm}=\frac{\omega \pm ku_d \sin \theta}{\sqrt{2}kv_{th}}$. Eigenmodes of the system correspond to solving $\text{det} (D)=0$ for $\omega$ with corresponding eigenvectors satisfying $R^TDR\vec{E}=0$, where $R$ is the matrix for rotation by angle $-\theta$. 

At any given $v_{th}$ and $u_d$, there will be a fastest-growing mode at some $k=|\vec{k}|$ and $\theta$. As seen in Figure~\ref{fig:angle_growth}, the fastest-growing mode transitions from $\theta=0^\circ$ to $\theta=90^\circ$ in the weakly relativistic region $0.1c\lesssim u_d\lesssim 0.5c$. The TS growth rate dominates that of the FI for $u_d \lesssim 0.1c$, because the FI is an electromagnetic instability, generally growing by a factor of $O(u_d/c)$ slower. While for $u_d> 0.5c$, the FI growth rate dominates; however, at these flow velocities relativistic effects become significant and the non-relativistic dispersion relation from equation~(\ref{eq:D}) provides only qualitative results.

In the transition region, most of the modes have growth rates roughly within a factor of 3-4 of each other. While the fastest-growing mode will reach saturation first, the other modes can continue to grow approximately at their linear rates, because the nonlinear saturation stage of the dominant mode may not have had enough time to significantly alter the property of the initial distribution function driving the instability. For example, TS saturation may scatter the distinct counter-streaming motions that drive the FI, but an effective temperature anisotropy capable of driving the secular WI may remain. This competition leads to interaction between the nonlinear stages of many of the modes, hence the final steady state can be different than simply the steady state of the fastest-growing mode alone. 

\section{Simulation Results}
To study the simultaneous competition of TS, Oblique, and FI modes, we initialize a bath of electric and magnetic fluctuations with random amplitudes and phases, e.g.,

\begin{equation}\label{eq:B}
B_z(t=0)=\sum_{n_x,n_y=0}^{16,16}\tilde B_{n_x,n_y}\sin(\frac{2\pi n_x x}{L_x}+\frac{2\pi n_y y}{L_y}+\tilde \phi_{n_x,n_y}).
\end{equation}
$E_x(t=0)$ and $E_y(t=0)$ are initialized similarly, and they are given equal average energy densities, $\langle  \epsilon_0 E_x^2/2\rangle=\langle \epsilon_0 E_y^2/2\rangle=\langle B_z^2/2\mu_0\rangle\approx 10^{-7}E_K$.

We present the results of four simulations, shown by the red crosses in Figure~\ref{fig:angle_growth}, where we fix the drift velocity at $u_d=0.1c$, but vary the temperature of the beams by choosing $v_{th}/u_d\in \{0.1,0.175,0.25,0.5\}$. The box sizes, respectively, are $L_x/d_e\in \{2.7,3.8,4.4,7.7\}$ and $L_y/d_e\in \{3.1,4.0,4.8,6.3\}$. Box sizes $L_x=2\pi/k_{FI}^{max}$ and $L_y=2\pi m/k_{TS}^{max}$ are chosen to be roughly equal $L_x\approx L_y$ while fitting a single fastest-growing wavelength of FI and an integer number, $m\approx k_{TS}^{max}/k_{FI}^{max}$, of TS wavelengths. All simulations used the Serendipity polynomial basis of order 2 for phase space discretization \citep{arnold11,juno18}. The evolution of the magnetic 

\begin{figure}[h]
\centering
\includegraphics[width=.9\linewidth]{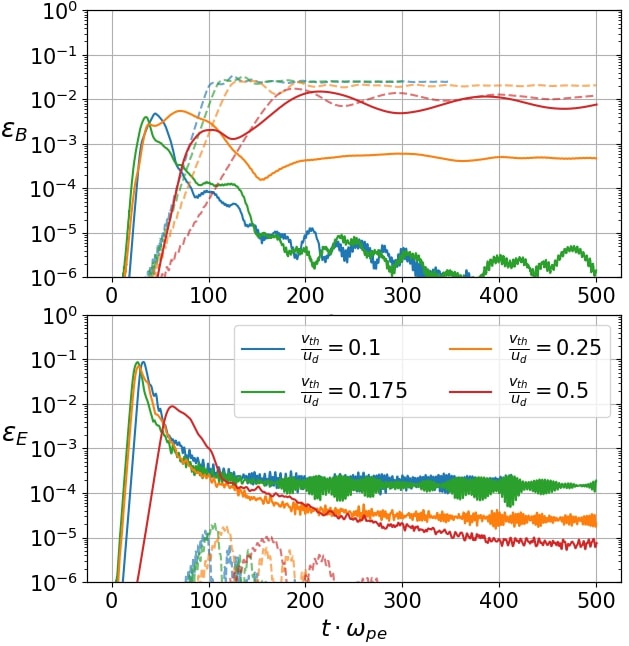}
\caption{ Growth and saturation of magnetic (top panel) and electric (bottom panel) energies normalized by the initial kinetic energy for beams with drift velocity $u_d=0.1c$ at different temperatures. Solid lines correspond to 2X2V with initial random modes, while dashed lines correspond to 1X2V with pure FI modes. }\label{fig:growth_sat}
\end{figure}
\begin{figure}[h]

\includegraphics[width=.972\linewidth]{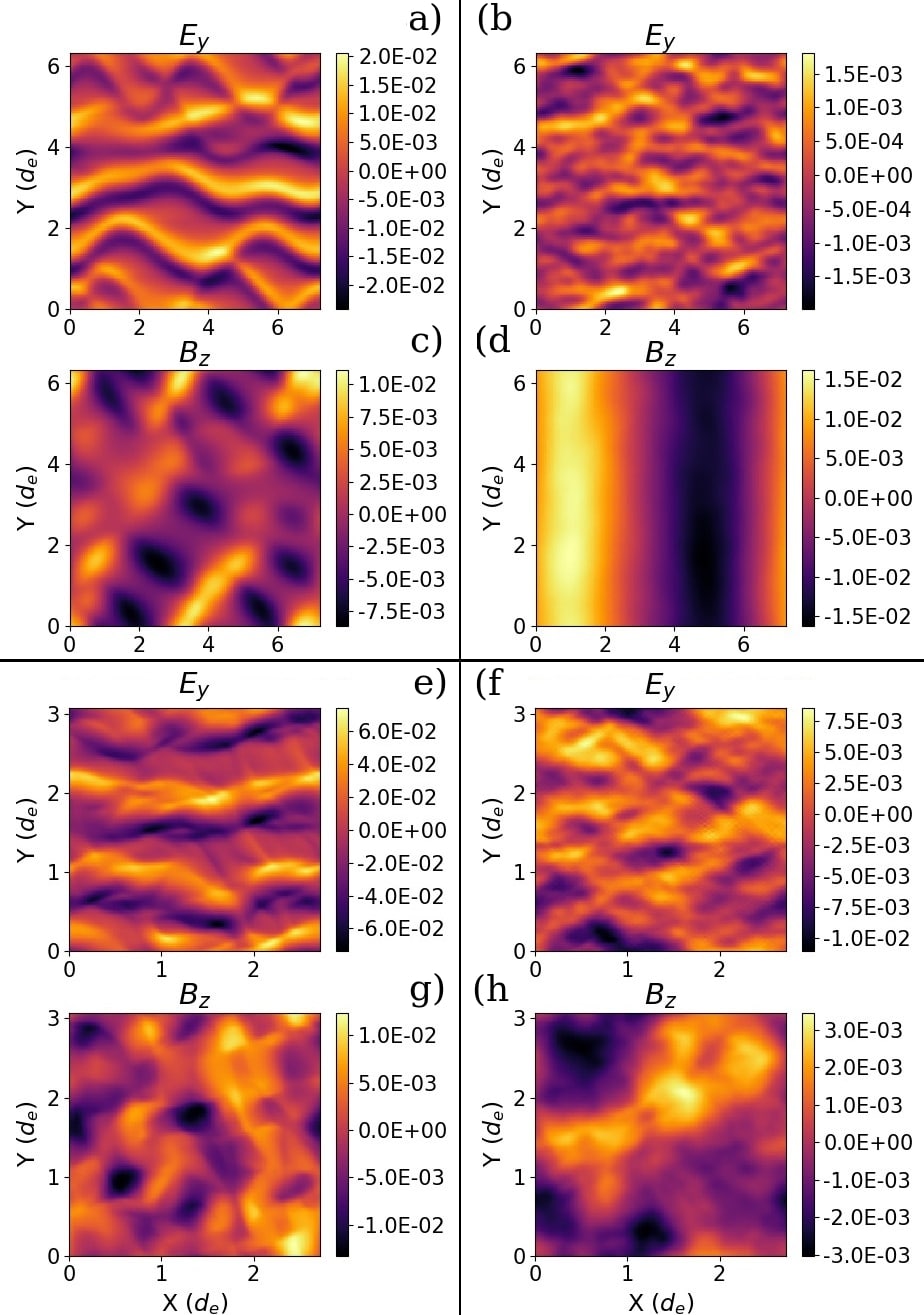}
\caption{Snapshots of electric, $E_y$, and magnetic, $B_z$, fields. The top and bottom four plots correspond to hot and cold cases, respectively. The left and right columns correspond to times during saturation ($t\omega_{pe}=35$ for cold and $t\omega_{pe}=80$ for hot case) and long after
($t\omega_{pe}\approx 250$ for both cold and hot cases).}\label{fig:fields}
\end{figure}
($\epsilon_B$) and electric ($\epsilon_E$) energy densities normalized to the initial kinetic energy is shown in Figure~\ref{fig:growth_sat} as solid lines for 2X2V with random initial modes and as dashed lines for 1X2V with pure 1X FI initial modes at the same beam parameters.

In all 2X2V simulations, a combination of TS and multiple Oblique modes exponentially grow the fastest during the linear phase, as predicted by the linear growth rates in Figure~\ref{fig:angle_growth}, and then are the first to saturate. The presence of electromagnetic Oblique modes is indicated in Figure~\ref{fig:growth_sat} by the exponential growth of both electric and magnetic energy, as pure TS modes would appear only as growing electric energy. The simultaneously present FI modes grow much more slowly, and during the linear phase are comparable to the pure FI modes in 1X2V shown as dashed lines in Figure~\ref{fig:growth_sat}. Following saturation, the nonlinear interaction of potential wells formed by the saturation of two-stream and Oblique modes, the tilted current filaments of Oblique modes, and the vertical current filament formation associated with the potentially still-growing FI become dominant. 

The nonlinear dynamics forming or disrupting current filaments are strongly temperature dependent. Figure~\ref{fig:fields} shows spatial snapshots of $E_y$ and $B_z$ for the (upper panels) hot, $v_{\mathrm{th}}/u_d=0.5$, and (lower panels) cold, $v_{\mathrm{th}}/u_d=0.1$, cases during and long after the time of saturation. In the post-saturation phase of the hot case, the horizontally varying $B_z$ associated with vertical current filaments is dominantly visible and steady in time (Figure~\ref{fig:fields}d), while in the cold case $B_z$ is spatially disorganized and strongly fluctuating in time (Figure~\ref{fig:fields}h). We find that a qualitatively similar pattern occurs for $u_d/c\in \{1/30, 0.25, 0.4\}$. Note that $u_d/c=1/30$ is the upper limit of intergalactic plasma drift velocities relevant to cosmological scenarios and cosmological magnetic field generation \citep{miniati02,schlick03,lazar09}. 

\begin{figure}[h]

\includegraphics[width=\linewidth]{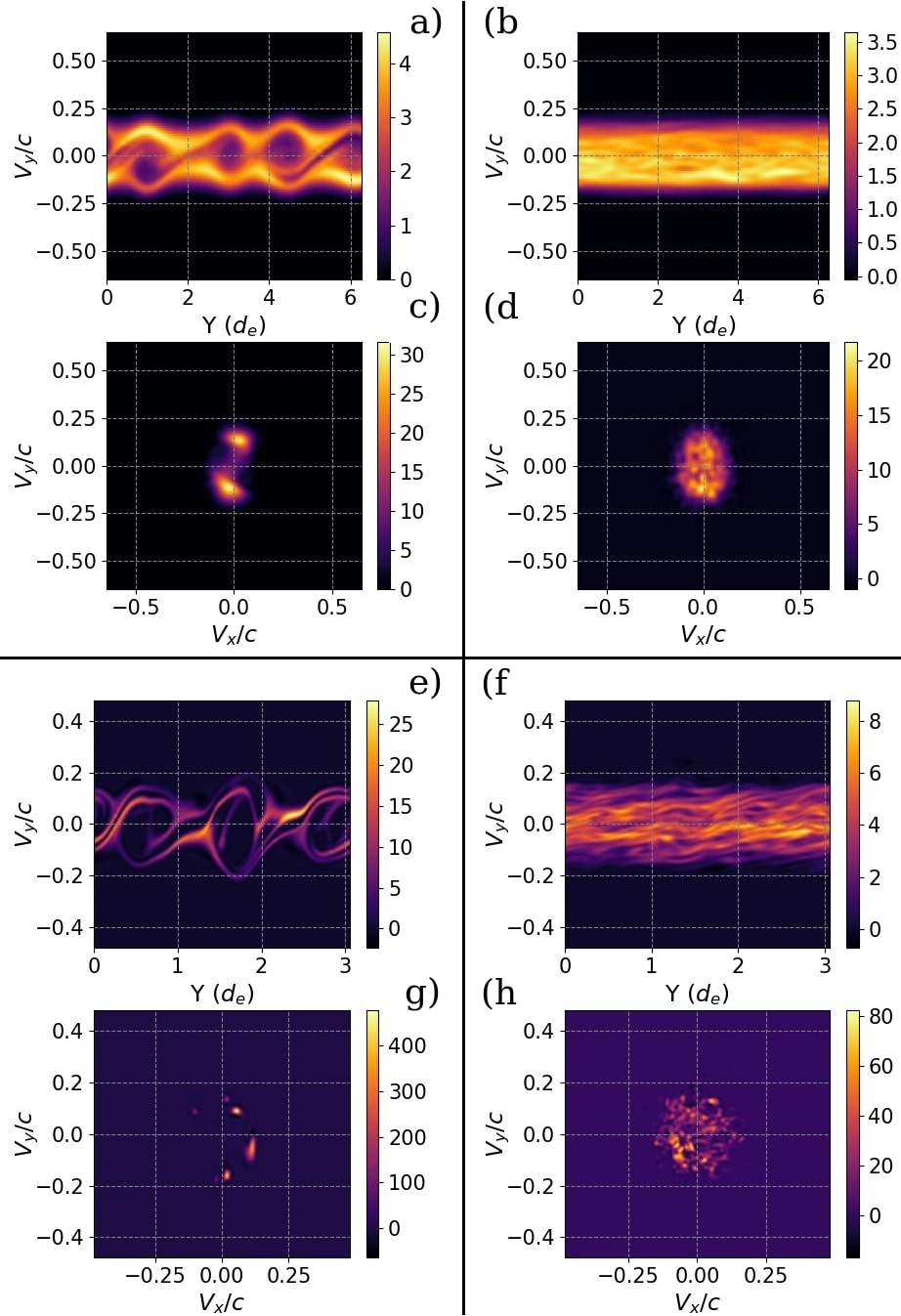}

\caption{(a), (b), (e), and (f) show the velocity distribution function in the $Y$ vs. $V_y$ plane at $X=L_x/2$ and integrated over $V_x.$  (c), (d), (g), and (h) show the velocity distribution function in the $V_x$ vs. $V_y$ plane plane at $(X,Y)=(L_x/2,L_y/2)$. The grouping, with respect to simulation parameters and time, is the same as in Figure~\ref{fig:fields}. }\label{fig:phase_space}
\end{figure}

\begin{figure}[h]

\includegraphics[width=\linewidth]{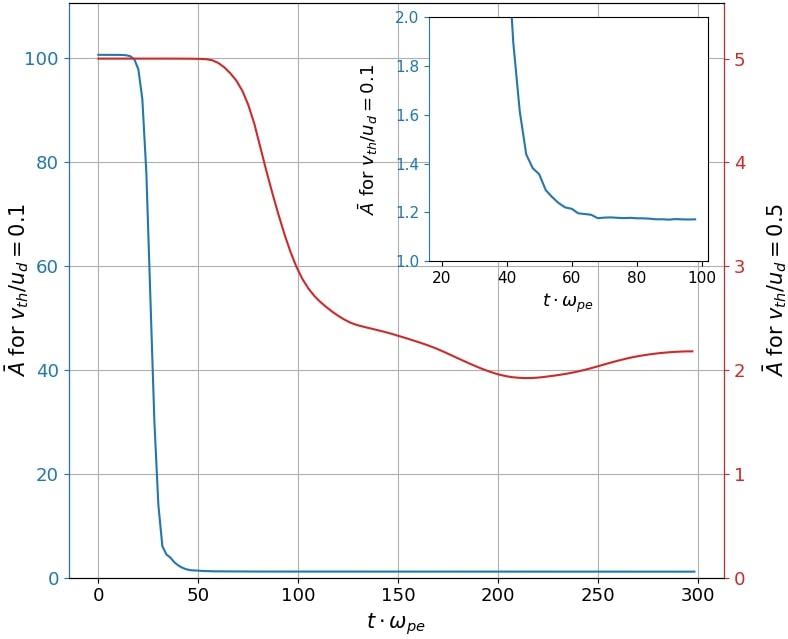}

\caption{Spatially averaged temperature anisotropy $\bar{A}$ vs. time for the cold case in blue (left axis) and the hot case in red (right axis). The inset shows a zoom-in of the the cold case, where $\bar{A}$ both drops and levels off sharply during saturation.}\label{fig:aniso}
\end{figure}

In all cases, the electric energy quickly grows, saturates, and then rapidly decreases. In Figure~\ref{fig:phase_space} we plot different parts of the phase space before and after the disruption of the electric energy for the same hot and cold cases shown in Figure~\ref{fig:fields}. In the hot case, near the time of saturation, clear 2D electron tubes form (Figure~\ref{fig:fields}a) due to the fastest-growing wavelength of the TS $y$-varying $E_y$ extending in the $x$ direction, which appear as 1X electron holes, or BGK modes, in the $Y$ versus $V_y$ cut at a fixed $X$ value shown in Figure~\ref{fig:phase_space}a. The 2D electron tubes are slightly skewed due to the simultaneously growing, albeit slower, Oblique modes. TS saturation creates a plateau in the $V_y$ direction of the velocity distribution function (Figure~\ref{fig:phase_space}d), leaving an effective temperature anisotropy from which the now-secular Weibel instability can grow. In Figure~\ref{fig:aniso}, we find that the spatially averaged temperature anisotropy, $\bar{A}$, drops from $\bar{A}=5$ at $t\omega_{pe}=40$ just before saturation to $\bar{A}\approx 3$ at $t\omega_{pe}=100$ soon after, where
\begin{equation}
    \bar{A}=\int_0^{L_y} \int_0^{L_x} \frac{\int (v_y-\bar{v}_y)^2f(x,y,\vec{v})\mathrm{d}\vec{v}}{\int (v_x-\bar{v}_x)^2f(x,y,\vec{v})\mathrm{d}\vec{v}}\,\mathrm{d}x\mathrm{d}y,
\end{equation}

and $\bar{v}_i=\int v_i f(x,y,\vec{v}) \mathrm{d}\vec{v}/\int f(x,y,\vec{v}) \mathrm{d}\vec{v}$. On the order of an electron bounce period, $T_b$, in the BGK mode after saturation, filaments from the previous FI and now growing secular WI disrupt all of the electron tubes, leading to the observed sudden decrease in stored, what is primarily, electrostatic energy, where $T_b \sim 2\times 2\pi/(k_{TS}^{max} u_d)\sim 20\omega_{pe}^{-1}$. The remaining temperature anisotropy of $\bar{A}\approx 3$ at $t\omega_{pe}=100$ drives further Weibel growth, and visible filaments appear around $t\omega_{pe}=150$ (not shown), where $\epsilon_B$ grows again. The magnetic energy then saturates around $\epsilon_B\sim 10^{-2}$ near the Alfv\'{e}n-limited regime, $\rho_g\sim m_e u_d/(e B_z)\sim 7d_e\sim L_x$, and enters a steady-state oscillation at the magnetic bounce frequency, agreeing closely with PIC and 1X2V simulations.

In the cold case, many Oblique modes with $\theta \gtrsim 30^\circ$ have the same or faster growth rates than TS and therefore saturate at similar times, creating 2D electron tubes with far more spatial structure and with tilted current filaments passing through (Figure~\ref{fig:fields}e and \ref{fig:phase_space}e). The phase space has sharp structures with the main two beams rotating in the $V_x-V_y$ plane (Figure~\ref{fig:phase_space}g), followed by the beams fragmenting and developing fine structure, as shown in the distribution functions in Figure~\ref{fig:phase_space}h and \ref{fig:phase_space}f at much later time. The spatially averaged temperature anisotropy sharply drops (Figure~\ref{fig:aniso}) from $\bar{A}=101$ at $t\omega_{pe}=20$ immediately prior to saturation, to $\bar{A}\approx 1.2$ at $t\omega_{pe}=50$ soon after. However, unlike the hot case, $\bar{A}$ remains constant through the rest of the simulation. Filaments do not form after disruption of the BGK modes due to the lack of a significant residual effective temperature anisotropy to drive the secular WI, and the end result is a bath of primarily electrostatic fluctuations. The lack of current filaments does not provide a steady structure to sustain magnetic fields, thus $\epsilon_B$ decays to values orders of magnitude smaller than that observed in PIC and 1X2V simulations.

We have doubled and tripled the box sizes and found no qualitative difference in the results. In the hot case, filaments form at the same wavelengths $\lambda_{FI}^{max}=2\pi/k_{FI}^{max}$ as before, and then merge to larger scales. However, the merging process does not break up the filaments as was reported in the 2X case of recent PIC simulations \citep{kumar15,takamoto18}, although their beam drift velocities were relativistic, with relativistic $\gamma=5$. In the cold case, we still see no filament formation for larger box sizes. This result contradicts previous non-relativistic PIC simulations of \citet{kato08}, in which they found filament formation and $\epsilon_B\sim 10^{-2}$ post-saturation for the same cold beam parameters. Note that in the cold case, our result is physically different than in \citet{kumar15} and \citet{takamoto18}; they see the disruption of filaments after the merging of initial filaments forming at $\lambda_{FI}^{max}$ scales, while we do not observe formation of the $\lambda_{FI}^{max}$ scale filaments in the first place. Inherent particle noise in PIC simulations may potentially make beams effectively hotter than they should be, which weakens the effects in the cold beam regime that diminish the effective temperature anisotropy.

To disentangle the role of TS and Oblique modes versus temperature anisotropy, we also ran 2X2V simulations of the classic Weibel instability with a bi-Maxwellian initial distribution function: 
\begin{equation}
f_{0,e}(v_x,v_y)=\frac{m_e}{2\pi\sqrt{ k_BT_x k_BT_y}}e^{-\frac{-m_ev_x^2}{2k_BTx}-\frac{m_ev_y^2}{2k_BTy}}, 
\end{equation}
with $k_BT_y=m_e\left(u_d^2+v_{th}^2\right)$ and $k_BT_x=m_ev_{th}^2$ for the same anisotropy parameters $\bar{A}=101$ and $\bar{A}=5$. Current filament formation and merging were unaffected by temperature anisotropy, achieving $\epsilon_B\sim 10^{-2}$ in the nonlinear steady state for both cases. This result emphasizes the role that TS and Oblique mode saturation play in affecting the effective temperature anisotropy of the distribution function and in controlling the growth of current filaments via the secular WI.

\section{Conclusions}
We present the first 2X2V continuum-Vlasov Maxwell simulations of weakly relativistic interpenetrating plasma flows to show a strong dependence of current filament formation and resulting magnetization on the internal temperature of the beams. Hotter beams maintain an effective temperature anisotropy after saturation of the faster growing TS and Oblique modes, allowing for continuing current filament formation via the secular Weibel instability. The result is a final magnetization of $\epsilon_B\sim 10^{-2}$ due to current filaments reaching the Alfv\'{e}n current limited regime and merging to larger scales. On the other hand, the effective temperature anisotropy for colder beams falls rapidly and remains constant after the saturation of TS and Oblique modes. Current filament formation does not follow, and instead, a bath of electrostatic fluctuations dominates with effectively no magnetization, $\epsilon_B\lesssim 10^{-5}$. 

These results put testable constraints on the efficiency of magnetic field generation via Weibel-type instabilities in astrophysical and laboratory scenarios. Differences in observed magnetization for colliding plasmas of similar flow velocities could potentially be attributed to a difference in the beams' internal temperatures. We have found this temperature dependence for the range $1/30\leq  u_d/c\leq 0.4$, where $u_d/c=1/30$ corresponds to the upper bound of intergalactic plasma counter-streaming velocities relevant to cosmological scenarios. The source of disagreement between previous PIC simulations, which find filament formation, and our study, which does not, remains to be determined in the case of colder beams. The continuum Vlasov-Maxwell calculations presented in this Letter could be extended to include the evolution of an ion species, as well as the third dimension, while PIC calculations could be repeated with larger numbers of particles-per-cell to further mitigate the noise endemic to the PIC algorithm, even beyond the standard convergence analyses that have been performed in previous studies.



\acknowledgments
We are grateful for insights from conversations with Petr Cagas and William Fox.
This work used the Extreme Science and Engineering Discovery Environment (XSEDE), which is supported by National Science Foundation grant No. ACI-1548562, resources of the Argonne Leadership Computing Facility, which is a DOE Office of Science User Facility supported under Contract DE-AC02-06CH11357, as well as the Eddy cluster at Princeton University. A. H. was supported by the U.S. Department of Energy under Contract No. DE-AC02-09CH11466 and Air Force Office of Scientific Research under grant No. FA9550-15-1-0193; J. J. was supported by a NASA Earth and Space Science Fellowship (Grant No. 80NSSC17K0428); and J.M. TenBarge was support by NSF SHINE award (AGS-1622306).

\clearpage


\begin{thebibliography}{}
\bibitem[Arnold \& Awanou (2011)]{arnold11}{ Arnold, D. N., \& Awanou, G. 2011, Foundations of Computational Mathematics, 11, 337 }
\bibitem[Bret(2009)]{bret09}Bret, A. 2009, ApJ, 699, 990
\bibitem[Cagas et al.(2017)]{cagas17} Cagas, P., Hakim, A., Scales, W., \& Srinivasan, B. 2017, PhPl, 24, 112116
\bibitem[Califano et al.(1998)]{calif98}Califano, F., Pegoraro, F., Bulanov, S. V., \& Mangeney, A. 1998, PhRvE, 57, 7048
\bibitem[Davidson et al.(1972)]{davidson72} Davidson, R. C., Hammer, D. A., Haber, I., \& Wagner, C. E. 1972, PhFl, 15, 317 

\bibitem[Fonseca et al.(2003)]{fonseca03}Fonseca, R. A., Silva Luı́s O., Tonge, J. W., Mori, W. B., \& Dawson, J. M. 2003, PhPl, 10, 1979
\bibitem[Fox et al.(2013)]{fox13}Fox, W., Fiksel, G., Bhattacharjee, A., et al. 2013, PhRvL, 111, 225002
\bibitem[Frederiksen et al.(2004)]{fred04}Frederiksen, J. T., Hededal, C. B., Haugbølle, T., \& Nordlund, Å. 2004, ApJL, 608, L13
\bibitem[Fried(1959)]{fried59}Fried, B. D. 1959, Physics of Fluids, 2, 337
\bibitem[Hededal et al.(2004)]{hededal04}Hededal, C. B., Haugbølle, T., Frederiksen, J. T., \& Nordlund, Å. 2004, ApJL, 617, L107 
\bibitem[Huntington et al.(2015)]{hunt15}Huntington, C. M., Fiuza, F., Ross, J. S., et al. 2015, Nature Physics, 11, 173
\bibitem[Juno et al.(2018)]{juno18}Juno, J., Hakim, A., Tenbarge, J., Shi, E., \& Dorland, W. 2018, JCoPh, 353, 110
\bibitem[Kato \& Takabe(2008)]{kato08}Kato, T. N., \& Takabe, H. 2008, ApJL, 681, L93
\bibitem[Kazimura at al.(1998)]{kazimura98}Kazimura, Y., Sakai, J. I., Neubert, T., \& Bulanov, S. V. 1998, ApJL, 498, L183
\bibitem[Kumar et al.(2015)]{kumar15}Kumar, R., Eichler, D., \& Gedalin, M. 2015, ApJ, 806, 165
\bibitem[Lazar et al.(2009)]{lazar09}Lazar, M., Schlickeiser, R., Wielebinski, R., \& Poedts, S. 2009, ApJ, 693, 1133
\bibitem[Medvedev \& Loeb(1999)]{medved99}Medvedev, M. V., \& Loeb, A. 1999, ApJ, 526, 697
\bibitem[Miniati(2002)]{miniati02}Miniati, F. 2002, MNRAS, 337, 199
\bibitem[Nishikawa et al.(2003)]{nishi03}Nishikawa, K.-I., Hardee, P., Richardson, G., Preece, R., Sol, H., \& Fishman, G. J. 2003, ApJ, 595, 555
\bibitem[Nishikawa et al.(2005)]{nishi05}Nishikawa, K.-I., Hardee, P., Richardson, G., Preece, R., Sol, H., \& Fishman, G. J. 2005, ApJ, 622, 927 
\bibitem[Sakai et al.(2004)]{sakai04}Sakai, J.-I., Schlickeiser, R., \& Shukla, P. 2004, PhLA, 330, 384
\bibitem[Schlickeiser \& Shukla(2003)]{schlick03}Schlickeiser, R., \& Shukla, P. K. 2003, ApJL, 599, L57
\bibitem[Silva et al.(2003)]{silva03}Silva, L. O., Fonseca, R. A., Tonge, J. W., et al. 2003, ApJL, 596, L121
\bibitem[Takamoto et al.(2018)]{takamoto18}Takamoto, M., Matsumoto, Y., \& Kato, T. N. 2018, ApJL, 860, L1
\bibitem[Weibel(1959)]{weibel59}Weibel, E. S. 1959, PhRvL, 2, 83


\end{thebibliography}
\end{document}